\documentstyle[preprint,aps,epsfig]{revtex}
\begin{document}
\def\lsim{\mathrel{
\def\arraystretch{.3}
\begin{array}{c}
$$<$$\\
$$\sim$$
\end{array}}}
\def\gsim{\mathrel{
\def\arraystretch{.3}
\begin{array}{c}
$$>$$\\
$$\sim$$
\end{array}}}
\draft

\title{Nanoscale self-affine surface smoothing by ion bombardment}

\author{D. K. Goswami and B. N. Dev$^*$}
\address{Institute of Physics, Sachivalaya Marg, Bhubaneswar 751005, India}
\maketitle

%\begin{center}
%\it{Abstract}
%\end{center}

\begin{abstract}
Topography of silicon surfaces irradiated by a 2 MeV Si$^+$ ion beam at
normal incidence and ion fluences in the range $10^{15}-10^{16}$ ions/cm$^{2}$
 has been investigated using scanning tunneling microscopy. At
length scales below $\sim$~50 nm, surface smoothing is observed; the smoothing
is more prominent at smaller length scales. The smoothed surface is self-affine
with a scaling exponent $\alpha=0.53\pm0.02$.\\
\end{abstract}


One of the fundamental problems in materials science is to understand
the effects of particle radiation on solid surfaces. The evolution of solid
surface topography during ion-beam irradiation is governed by the interplay 
between the dynamics of surface roughening due to sputtering and smoothing due 
to material 
transport during surface diffusion. These competing processes are responsible 
for the creation of characteristic surface features like quasiperiodic 
ripples\cite{chason,carter,wittmaack,habenicht} and self-affine 
topographies\cite{habenicht,eklund,krim}. These have been observed in the ion energy
regime where sputtering is dominant and ion incidence is tilted to the surface
normal.  Although there is a large number of 
observations of ripple formation there are only a few studies on the 
scaling of the surfaces evolved in ion bombardment\cite{habenicht,eklund,krim}. 
A common feature of 
most rough surfaces observed experimentally or in discrete
models is that their roughness follows simple scaling laws. Surface 
root-mean-square roughness $\sigma$ is defined as 
$\sigma = <[h(x,y) - h]^2>^{1/2}$,
where $h(x,y)$ is the surface height at a point $(x,y)$ on the surface 
and $h$ is the 
average height. The surface is termed self-affine if $\sigma$ changes with the
horizontal sampling length $L$ according to $\sigma~~ \infty ~~L^\alpha$, 
where  $0 < \alpha < 1$ is the roughness exponent\cite{krim}. The roughness 
exponent quantifies how roughness changes with length scale and its value 
is indicative of the surface texture. \\

For graphite bombarded with 5 keV Ar ions at an angle $\theta = 60^\circ$ 
with respect to the surface normal, Eklund et al\cite{eklund}. reported $\alpha 
\simeq 0.2-0.4$, consistent with the predictions of the Kardar-Parisi-Zhang 
(KPZ) equation in 2+1 dimensions. Krim et al\cite{krim} observed
 a self-affine surface roughness generated by 5 keV Ar ion 
bombardment of an Fe thin film 
sample at $\theta = 25^\circ$, with a scaling exponent $\alpha$=0.53, 
with no theoretical model predicting this value. In all these cases an increase 
of surface roughness was observed due to ion bombardment.  Since ion 
arrival  on the surface is a stochastic process and 
sputtering events are spatially distributed and of variable magnitude, surfaces 
are generally roughened during bombardment. In all the studies 
mentioned above the conditions are such that the erosion of the surface 
due to sputtering in ion bombardment is dominant
 over surface atomic diffusion. However, if the surface atomic diffusion 
dominates over sputtering, surface smoothing rather than roughening can 
occur\cite{carter}. Carter and Vishnyakov\cite{carter} 
have shown that inclusion of a 
directed flux of atoms parallel to the surface, generated by ion bombardment, 
in a stochastic differential equation description of the dynamics of 
surface evolution during sputter-erosion can induce
smoothing for near-normal ($\theta \approx 0$) ion incidence. The flux of 
atoms parallel to the surface provides an effective diffusion causing surface 
smoothing which competes with the roughening caused by sputtering. 
For $\theta \approx 0$, roughening is weak as sputtering yield is small 
and smoothing 
dominates. Indeed for an ion incidence angle $\theta \approx 0$,
 surface smoothing have been observed in ion bombardment over a large
 range of ion energies\cite{carter,gutzmann}. Although some observations 
of surface smoothing
 have been reported, to our knowledge there has been
 no scaling studies of ion-beam induced surface 
smoothing. In scaling studies 
for nonequilibrium film growth by deposition, a value of $\alpha\approx0.35$
is expected when surface mobility of deposited particles are not allowed and 
$\alpha$=0.66 is expected when surface mobility is 
allowed\cite{kardar,wolf,lai}.  For ion-induced roughening
the observed value of $\alpha$=0.2$-$0.4 is in reasonable agreement with the exponent for growth without surface diffusion. For ion-beam induced smoothing, 
where surface diffusivity is important,
one may expect a different value of the scaling exponent
 $\alpha$. \\

In this 
Letter, we present scanning tunneling microscopy (STM) characterization of 
surface smoothing in 2 MeV Si$^+$ ion irradiation of Si surfaces 
at normal incidence ($\theta = 0$). At length
scales below $\sim 50$ nm we observe smoothing of the ion-bombarded 
surface. The observed
value of the roughness exponent $\alpha =0.53 \pm 0.02$ indicates the 
self-affine nature of the smoothed surface. The ion irradiated surface shows 
smoother surface texture at smaller length scales. 
We have chosen MeV ions for which sputtering yield is small. In comparison, the 
collision-induced atomic displacement and effective surface diffusivity is
 large. Together with normal incidence, these conditions are 
expected to cause smoothing. The observation of scale dependent smoothing
with increased smoothing at smaller length scales
 has direct bearing on ion beam processing of nanostructures. \\

Si(100) substrates were irradiated with 2.0 MeV Si$^{+}$ ions in the ion 
implantation beam line of our 3 MV  tandem Pelletron 
accelerator\cite{ksekar,gkuri}. The ion beam
was incident along the surface normal ($\theta\approx0$) and rastered on 
the sample 
in order to obtain a uniformly irradiated area. 
One half of the sample was masked
and hence unirradiated. An ion beam flux of $\approx1\times10^{12}$ cm$^{-2}$ 
sec$^{-1}$ was used with  fluences in the range $10^{15}$ to
$10^{16}$ ions/cm$^{2}$. The
samples were kept at room temperature during ion irradiation. The pressure in 
the chamber was $\sim 10^{-7}$ mbar. The sample was then taken out of the 
irradiation chamber and inserted into a STM chamber (pressure: 
3$\times10^{-10}$ mbar) with an Omicron
variable temperature STM operating at room temperature. STM height calibration was done by measuring atomic step heights on clean Si(111)
and Si(100) surfaces. Roughness measurements were made on the pristine and the
irradiated halves of the sample. We did not remove the thin ($\sim$1.5 nm) 
native
oxide from the Si surface because the surface topography may be perturbed by
the effect of Ehrlich-Schwoebel barriers in different crystallographic
directions on a crystalline surface. In this regard the presence of the thin 
oxide layer is helpful and the effect of  the anisotropic diffusion
can be neglected.\\

In order to determine the roughness exponent from STM images we follow 
the procedure described in ref. 6. Typical STM images from the pristine 
and the irradiated (fluence $4\times10^{15}$ ions/cm$^2$) parts of a sample
are shown in Fig.1. A large number of scans, each of size $L$, were recorded
on the surface at random locations. The $\sigma$ values for the rms roughness
given by the instrument for the individual scans were then averaged. This 
procedure was repeated for many different sizes and a set of average $\sigma$
vs. $L$ values was obtained (each $\bar{\sigma}$ is the average of six to 
fifteen measurements). Each ${\sigma}$ value was computed after the 
instrument plane fitting and subtraction procedure had been carried out. 
$\bar{\sigma}$ vs. $L$ log-log plots for both halves of the sample are shown 
in Fig.2. For the ion-bombarded area of the sample we observe surface 
smoothing and by fitting the linear part of the data we
 obtain $\alpha = 0.53\pm0.02$ below a length scale of 
$\simeq$ 50 nm, indicating the self-affine nature of the irradiated surface.
Below this length scale, the pristine half of the sample shows no linear 
region in the log-log plot of $\bar{\sigma}$ vs. $L$. Two vertical profiles
$h(x)$ along the lines marked in Fig.1. are shown in the inset of Fig.2. It
is also clear from these profiles that for the irradiated part of the 
sample the surface
is much smoother at shorter length scales as indicated by the roughness data
and the scaling exponent.\\

In earlier scaling studies\cite{eklund,krim} on ion-bombarded surfaces, 
the conditions of ion energy and the angle of incidence were favorable 
for strong sputtering and sputter-erosion of surfaces caused roughening.  
In order to explain the dominance of smoothing over roughening in our 
case let us first compare the sputtering yields. From the conditions 
in refs.[5] and [6], we estimate the sputtering yields of 3.7 atoms/ion 
and 3.9 atoms/ion, respectively, using the TRIM (transport of ions through 
matter) code\cite{trim}. In our case the higher ion energy and the normal 
incidence $-$ both contribute to lowering the sputtering yield, which is
$<$ 0.2 atom/ion. Thus the sputtering yield is smaller by almost a factor
 of 20. This indicates why surface erosion, main reason for 
roughness enhancement, is not significant in our case. In fact at large 
length scales surface roughness remains unaffected by ion bombardment.  
On the other hand, number of surface atoms that would contribute to 
effective surface mobility is large as discussed below.  In ion-atom 
collisions in solids and at the surface, the elastic energy lost by an 
ion is transferred to a recoil atom, which itself collides with other 
atoms in the solid and so forth. In this way the ion creates what is 
called a collision cascade. The displaced atoms in this collision cascade
 may acquire a kinetic 
energy enough to escape from the solid surface $-$ a phenomenon known as 
sputtering. However, if the energy (component normal to surface) 
of the displaced atoms is smaller than 
the surface binding energy, the atoms may reach the surface but cannot 
leave the surface. They can however drift parallel to the surface.
 We show the results of a TRIM simulation of sputtering 
yield for our case in Fig.3. This shows the atoms reaching the surface vs. 
their energies normal to the surface. Atoms which have energies greater
 than the surface binding energy ($\approx$ 4.7 eV) will be sputtered. 
However, we notice that a large number of atoms reach the surface with 
low energy ($<$ 4.7 eV) with the number of atoms/eV peaking at $\sim$ 1 eV.  
These atoms will not leave the surface (not be sputtered)\cite{*}. 
The role of these atoms is important in surface smoothing. These atoms 
have too low an energy (normal to surface) to escape the energy barrier 
at the surface and will translate parallel to the surface.  
This collision-induced atomic displacement and the consequent effective 
diffusivity parallel to the surface due to ballistic atomic transport 
can be the dominant surface relaxation
 mechanism.  As a result smoothing may dominate roughening as discussed 
later in more details. Eklund at al\cite{eklund} studied submicron-scale 
surface roughening induced by ion bombardment and obtained an scaling 
exponent $\alpha\simeq0.2 - 0.4$. This value of the exponent is reasonably 
explained by the anisotropic KPZ equation ($\alpha=0.38$)\cite{cuerno} 
when the surface diffusion term is expected to contribute negligibly. On 
the other hand, there are no concrete predictions of the exponents for 
the case where ion beam induced surface smoothing or diffusivity is dominant. 
Neither we know any scaling theory which predicts $\alpha\approx 0.5$. 
Assuming the possibility that the scaling theories applicable to 
nonequilibrium film growth may also be applicable to ion bombardment, 
so long as no eroded material is redeposited onto the surface, we compare 
the observed exponent with those expected for the deposition process, 
which are $\alpha\approx 0.35$ when surface mobility of the deposited
 particles is ignored and $\alpha=0.66$ when surface mobility is
 allowed\cite{kardar,wolf,lai}.  In the first case the exponents are in good 
agreement for deposition and ion bombardment.  In our case surface 
mobility is important and the observed  value of $\alpha=0.53$ is closer to
that for the deposition model that includes surface mobility. 
Incidentally, Krim {\it et al.}\cite{krim} also observed $\alpha=0.53$ for 
ion bombardment of an Fe film on a MgO substrate where roughening, rather 
than smoothing, was dominant.\\

For  ion irradiation, Carter and Vishnyakov\cite{carter} derived an 
expression showing the relative magnitudes of the roughening (sputtering) 
term and the smoothing term due to recoiled atoms which qualitatively 
explains the domination of smoothing over roughening at normal and near-normal 
($\theta \approx 0$) incidence of the ion beam. However, there is no prediction 
of scaling exponent. For $\theta\approx0$ they predict that smoothing 
dominates roughening at all wave vectors. We find that at larger length 
scales ($> 50$ nm) initial surface roughness remains practically unaffected 
by ion bombardment while smoothing becomes increasingly dominant at lower  
length scales below 50 nm.\\

In order to show the relative strength of the smoothing and the roughening 
terms, Carter and Vishnyakov\cite{carter} extended the treatment given by 
Bradley and Harper\cite{bradley}, who showed (in 1+1 dimension) that, due 
to sputter-erosion alone, the deterministic defining equation for $h(x,t)$ 
can be written as 
\begin{eqnarray}
-\frac{\partial h}{\partial t}=&& \frac{J}{N}~Y_0(\theta)-\frac{J}{N}
\frac{\partial}{\partial\theta}~\left[~Y_0(\theta)~cos~\theta~\right]
~\frac{\partial h}{\partial x}\nonumber\\
&&+ \frac{Ja}{N}~Y_0(\theta)~\Gamma_1(\theta)~\frac{\partial^2h}
{\partial x^2}
\end{eqnarray}
where $J$ is the mean ion flux incident at angle $\theta, N$ is the solid 
atomic density, $Y_{0}(\theta)$ is the sputtering yield of a plane surface,
$a$ is the mean depth of energy deposition by an ion, and $\Gamma_1(\theta)$ 
is a 
function of $\theta$, and standard deviations $\alpha$ and $\beta$ of the 
bi-Gaussian ellipsoidal ion energy spatial deposition density function. For 
order of magnitude estimation the ellipsoidal distribution has been
 approximated by a spherical distribution with $a=\alpha=\beta$, in which case
\begin{equation}
\Gamma_1(\theta) = sin^2~\theta - \left(cos^2~\theta/2\right)\left(1 + 
sin^2~\theta\right)
\end{equation}
In order to introduce the effective diffusion parallel to the surface they 
estimated the atomic flux parallel to the surface to modify the last 
 term in Eq.(1):
\begin{equation}
-\frac{J}{N}\left\{f(E) ~d~cos2\theta~-Y_0(\theta)~a~\left[sin^2~\theta 
-\frac{cos^2\theta}{2}(1 + sin^2\theta~)\right]\right\}
\frac{\partial~^2h}{\partial~x^2}
\end{equation}

\noindent where $f(E)$ is the no. of recoil atoms each ion generates in 
the solid and $d$ is the average distance traveled by the recoiled atoms. 
$d$ is of the order of a few interatomic distances. For $\theta=0$, the 
expression (3) [ i.e, the last term in Eq(1) ] is negative and smoothing 
dominates roughening at all wave vectors. $f(E) =k(E)/2E_d$, where $k(E)$ 
is the fraction of ion energy deposited in elastic collisions and $E_d$ is 
a displacement energy\cite{+}. In the simulation results shown in Fig.3 we 
have used $E_d$=15 eV. The results shown in Fig.3 only qualitatively shows 
how a large number of hyperthermal recoil atoms, arriving at the surface but 
unable to escape the surface, can cause surface smoothing as implied by Eq.(1) 
along with expression (3).  Expression (3) only qualitatively describes the 
effect of $f(E)$ in surface smoothing. For a quantitative understanding 
future theoretical work should include the effect of a distribution like 
that shown in Fig.3. So far theoretical works concentrated only on the low 
ion energy regime where sputter-erosion is dominant and the approximation 
($a=\alpha=\beta$) used in deriving the expression for $\Gamma_1(\theta)$ 
[ Eq.(2) ] may be valid. However, for high ion energies, it is not valid. 
For example in our case, for 2 MeV Si ions in Si, $a = 1.94$ $\mu$m, 
$\alpha=248$ nm and $\beta=288$ nm. In the existing theories it is assumed 
that energy released by the ions at a depth $a$ contributes an amount of 
energy to surface points that may induce surface atoms to break their bonds 
and leave the surface\cite{cuerno}. This is true for low ion energies where 
$a$ is small. However, ion energy release deep inside the sample would hardly
have any effect on surface atoms. Future theories must take this aspect into
account.\\

In conclusion, we have observed nanoscale surface smoothing in ion bombardment. 
The smoothed surface is a self-affine fractal surface with a scaling exponent 
$\alpha=0.53\pm0.02$. Below a length scale of $\sim50$ nm, the smoothing is 
more dominant at smaller length scales. This phenomenon may be used in 
reducing surface roughness of nanostructural devices by ion beam processing 
as ion beams are widely used in device fabrication. Transport in nanostructures
is expected to improve when roughness is minimized.  For an understanding of 
the scaling exponent observed in surface smoothing further theoretical studies 
will be necessary.

\noindent $^*$Corresponding author: bhupen@iopb.res.in

\newpage

{\bf {FIGURE CAPTIONS:}}\\

Fig.1. STM images recorded on a  pristine (top) and ion-bombarded (bottom)
silicon surface. The scan size is 300 $\times$ 300 nm$^2$ and the vertical
scale (black to white) is 2.2 nm. Height  profiles along the lines are 
shown in Fig.2.\\

Fig.2. Average root-mean-square roughness vs. scan size on the pristine and the 
ion irradiated surfaces. Each point represents an average of 6 to 15 scans 
recorded at random locations on the surface. 
Surface smoothing is observed at scan sizes below $\sim 50 \times 50$ nm$^2$. 
The least-squares fit (solid line) to the linear portion of the data for the 
irradiated sample gives the
scaling exponent $\alpha = 0.53\pm0.02$. No linear part is observed for
the pristine sample data. Two vertical profiles 
$h(x)$ measured along the lines marked in Fig.1, are shown in the inset 
(scales in nm): (a) pristine, (b) irradiated sample.\\

Fig.3. A Monte-Carlo simulation result showing the energy distribution of ion-beam induced displaced atoms reaching the surface. Atoms with energy $>4.7$ eV leave the surface (sputtered). The large number of atoms below 4.7 eV (surface binding energy) cannot leave the surface and contribute to an effective surface diffusion due to ballistic atomic transport leading to smoothing.

\begin{center}
\begin{figure}
Fig.1 (top)

\vspace*{1in}

\hspace*{-0.5cm}\includegraphics[height=13cm]{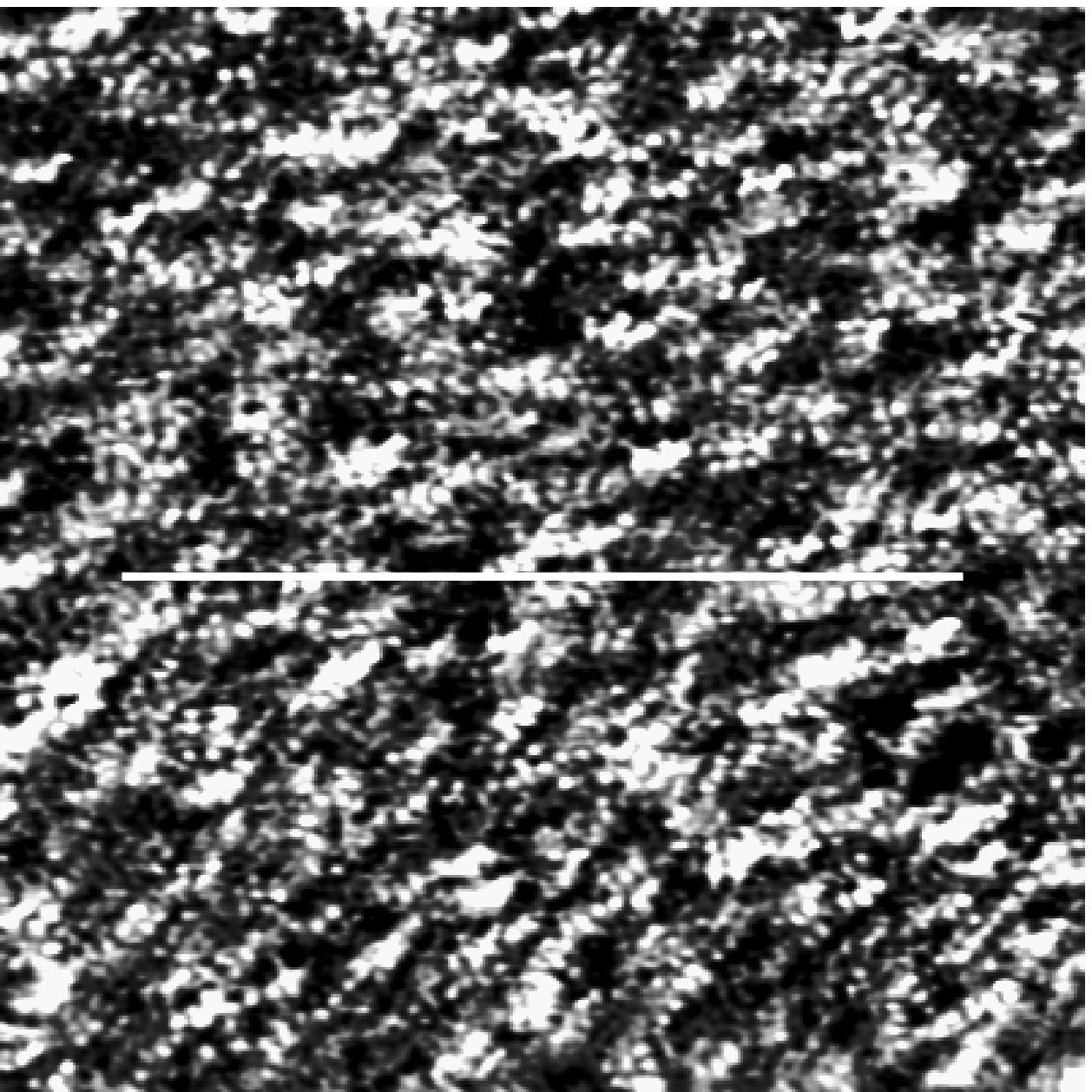}
\end{figure}
\end{center}
\newpage
\begin{center}
Fig.1 (bottom)
\begin{figure}
\vspace*{2in}
\hspace*{-0.5cm}\includegraphics[height=13cm]{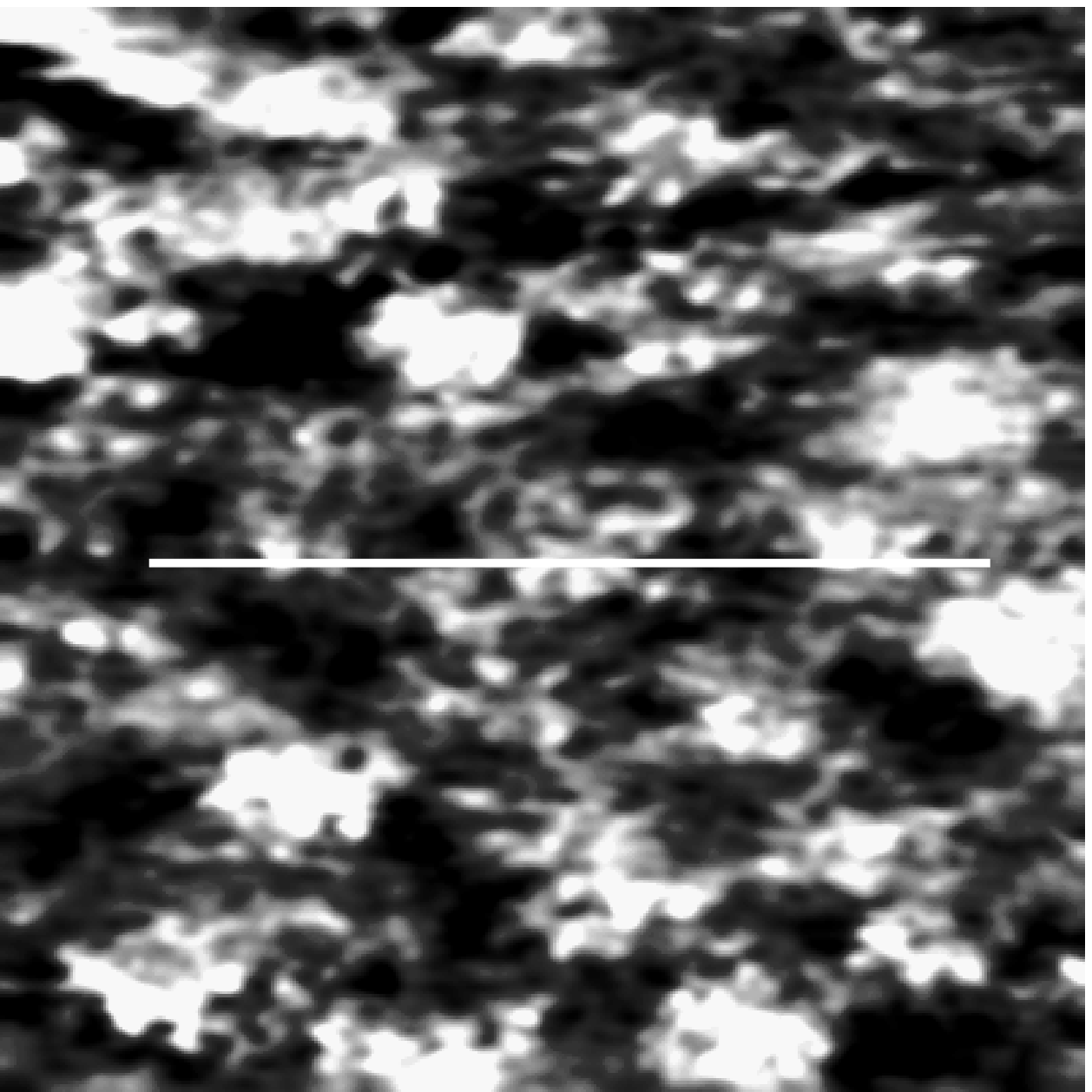}
\end{figure}
\end{center}
\newpage
\begin{center}
Fig.2.
\begin{figure}
\vspace*{2in}
\hspace*{-0.5cm}\includegraphics[height=13cm]{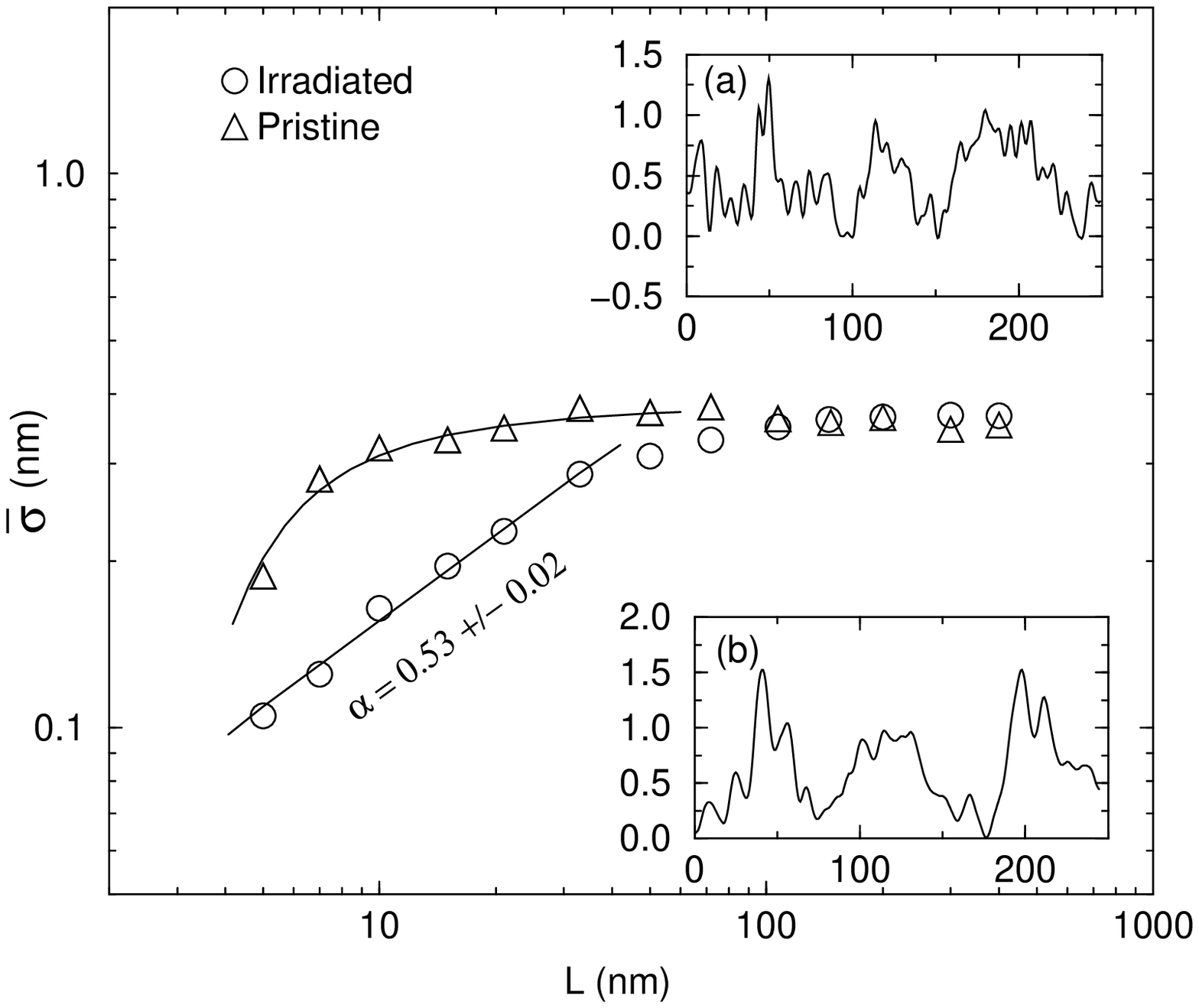}
\end{figure}
\end{center}
\newpage
\begin{center}
Fig.3
\begin{figure}
\vspace*{2in}
\hspace*{-0.5cm}\includegraphics[height=13cm]{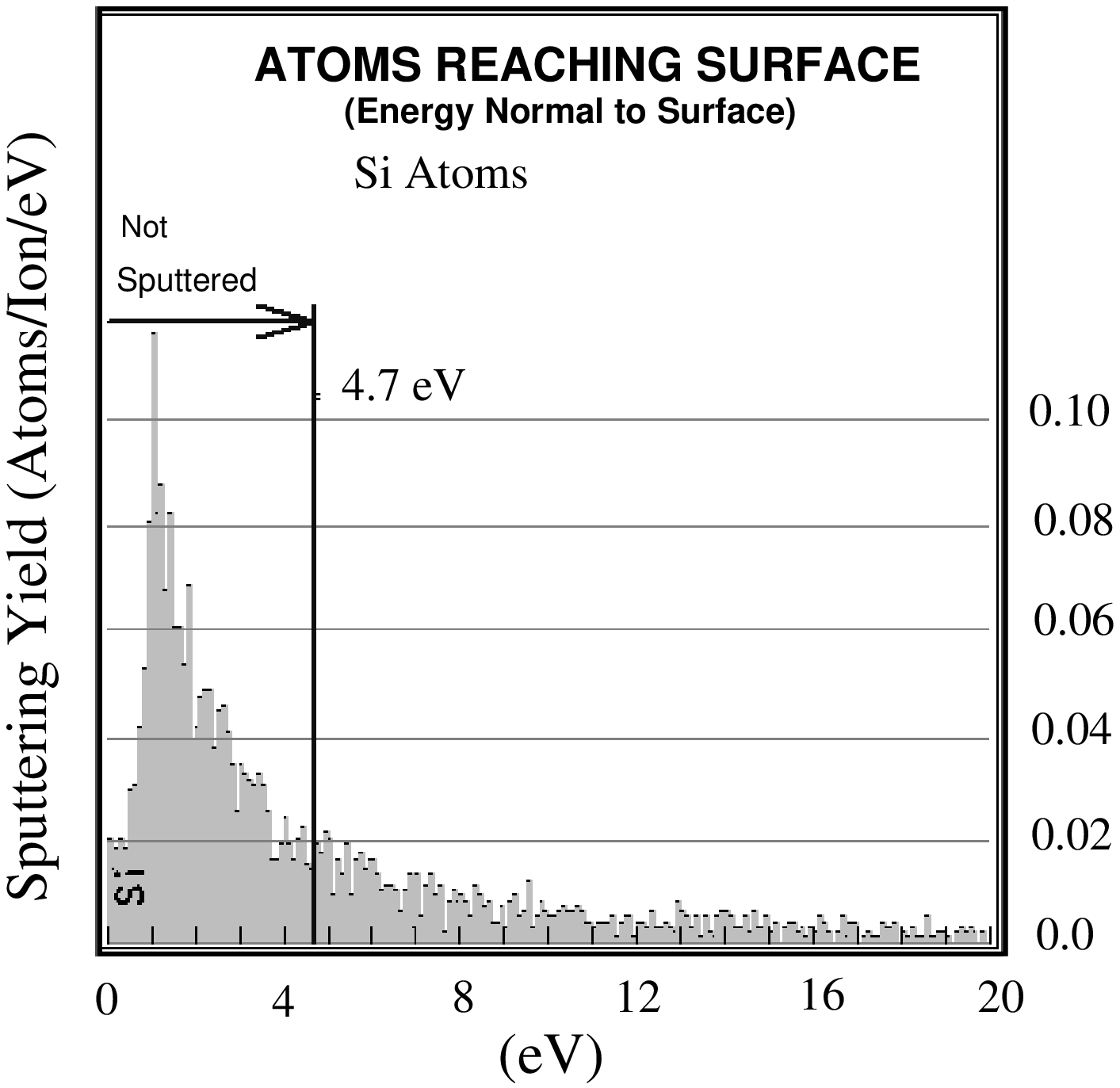}
\end{figure}
\end{center}

\begin{thebibliography}{99}
\bibitem{*}These qualitative nature remains the same when the native oxide layer
on the silicon surface is explicitly taken in the simulation. Both Si and O
atom distributions peak at $\sim$1 eV, O peak appearing at a slightly higher energy.
\end{thebibliography}
\end{document}